# The Definition and Long-term Variations of Beijing Blue Days

*Ruomu GAO [1], Yichuan Huang [2], Su Wang[3,4]*

1 Han Academy, Hong Kong
2 Capital Normal University High School
3 State Key Laboratory of Numerical Modeling for Atmospheric Sciences and Geophysical Fluid Dynamics, Institute of Atmospheric Physics, Chinese Academy of Sciences, Beijing 100029, China
4 University of Chinese Academy of Sciences, Beijing 100049, China

Corresponding author: Su Wnag (wangsu@mail.iap.ac.cn);

**Abstract**

The phenomenon of Beijing Blue days first occurs on June 11, 2015, and become a hot society topic in a short time. Thus, Beijing's blue day is not only what ordinary people desire but national needs. In this work, using the data of daily average meteorological observations during 1980 and 2014, we select three standards for defining Beijing blue day (BBI): no rain, clear sky（low cloud cover<=AX30）and great dry visibility（>=15km), whose accuracy is 73.4%. Our study finds that Beijing shares the most significant acceleration in the BTH region, with a growth rate of 2.66 d/year. The greatest seasonal average and the maximum acceleration rate appear in winter for total BBI and continuous BBI days, the least is in summer, while Beijing blue appears most frequently in January. BBI is more related to relative humidity and rainless day on an annual scale, but different in different seasons.

## 1.Introduction

Beijing, being the national political, scientific, and technological innovation center of China, has attracted the attention of the world for a period for its serious haze pollution. Haze is fine dust, smoke, or light vapor causing non-transparency of the air. The hazardous dust in the sky can lead to severe chronic disease and inhibited plant growth(Chen et al., 2015 , Cai et al., 2017). Moreover, by blurring the visibility of the sky and reducing the clarity to less than 10km, haze increases the probability of traffic accidents(Ramanathan et al., 2005). Therefore, there has been a lot of researches about the definition, causes, and impact of haze(Ramanathan et al., 2005 , Zhang et al., 2012 , Shen et al., 2015 , Su et al., 2015 , Zhang et al., 2016 , Liu et al., 2017b). In contrast, blue days, which occur frequently, are still easily ignored by researchers.

Up until now, with the development of the internet, the phenomenon of Beijing Blue Days had received a lot of attention. And it is not surprising that 'Beijing Blue' – close to people's daily life and essential to their wealth-- has become one of the ten clean air keywords published by Beijing Environmental Protection Publicity Centre in 2015. In recent years, there have been some researchers discussing the phenomenon of a temporary blue sky event appeared in Beijing, such as the 'Olympic Blue' during the 2008 Olympic game(Streets et al., 2007 , Lin et al., 2009 , Wang et al., 2009 , Wang et al., 2010 , Shen et al., 2011), 'APEC Blue' during the 2014 APEC meeting(Huang et al., 2015 , Guo et al., 2016 , Gao et al., 2017 , Li et al., 2017 , Liu et al., 2017a) , and 'Parade Blue' during the 2015 parade in Beijing(Kang et al., 2016 , Li et al., 2016 , Liu et al., 2016 , Wang et al., 2017 , Xue et al., 2018).

Meanwhile, meteorologists and ecologists define 'blue sky' differently, but neither of them matches the expectation of people. so, there have been still no clear definition and no long-term study on Beijing blue sky yet. Also, the influencing factors of the blue sky remain unknown. But recently, as the Chinese government had established policies to protect the blue sky in China in the 18th and 19th CPC National Congress(State Council of China 2018), establishing a definition of 'Beijing Blue' becomes necessary, so this study is useful for evaluating the achievement of these operations.

Following the Chinese blue days index developed by the work of Wang et al. (2019), we try to define Beijing blue index (BBI), and understanding the Spatio-temporal distributions of Beijing blue days shows indispensable important. for one thing, it provides a source of data for determining the period that is suitable to hold important major events in the capital, such as Asian Games. For another, it can help us control pollution, to get a better energy-saving emission reduction inventory. Moreover, it is more beneficial to the formulation of national policy, the choice of event events, the livable city, the list of energy-saving the sustainable development, and the promotion of the development of small and medium-sized cities.

## 2. Data and methods

After strict quality control(Lin et al., 2014 , Wang et al., 2019), daily average meteorological observations during 1980–2014 obtained from the National Meteorological Information Center (http://data.cma.cn/.) in Beijing station are used, which including temperature (T), precipitation (P), sunshine duration (SD), relative humidity (RH), wind speed (WS), total cloud cover (TCC), low cloud cover (LCC) and visibility (VIS) (observed four times per day at 2:00, 8:00, 14:00, 20:00 BJT).

Moreover, the hourly air quality index (AQI) in 2013 and 2014 from the China National Environmental Monitoring Center (Zheng et al 2014; http:// www.cnemc.cn/) are used to examine whether the BBI defined in this study can accurately represent the great air quality days variations in Beijing.

Furthermore, lacking weather phenomena records from the National Meteorological Information Center, we choose daily weather phenomena recorded by weather websites(http://www.tianqihoubao.com/lishi/index.htm) during 2011-2014 to examine whether the CBDI defined in this study can accurately represent the good air condition.

In this study, the correlation coefficient is used to reflect the relationship between two variables, it is defined as

$$r_{xt} = \frac{\sum_{i=1}^{n}(x_i - \overline{x})(y_i - \overline{y})}{\sqrt{\sum_{i=1}^{n}(x_i - \overline{x})^2 \sum_{i=1}^{n}(y_i - \overline{y})^2}},$$

in which $\overline{x}$, $\overline{y}$ is the mean value of the time series, n is the sample length. Also, the student's t-test is used to examine the significance of linear regression at the 95% confidence level.

**3. CBDI index definition and validation**

Wang et al. (2019) identify a sunny day as a day with daily precipitation ≤0.1 mm and low cloud cover≤75th percentile, but the choices depend on a national average range and may not suitable for Beijing. In our work, to get the most accurate definition for BBI, the definition procedure is divided into two steps: First, we scombine historical records with clear days (no rain and great visibility days, also called sunny days) and clean days (good air quality days with clean air ) to get the BBI; Second, we test the alternative definitions with a random experiment about the accuracy. Then we can find a suitable definition. The details for the definition are shown as follows:

(1) Firstly, we exclude days with precipitation. We select the reported important blue event related to Beijing and compare their meteorological factors on that day, the results are shown in Table 1. According to regulations, the sky can be divided into the sunny sky, light sky, broken sky and cloudy sky with clod fraction less than

10%, more than 10% and less than 30%, more than 30% and less than 70%, more than 70% respectively. And most investigations use the absolute index and total cloud to define a sunny day, but after we checked the blue event in historical reports and find all of them are in a no rainy day with low clod fraction less than 30% (AX30), but the total clod cover could be changed from 0% to 70%. Meanwhile, we also concern about percentile index (historical data corresponding to 70%, 75% of the low cloud cover: TX70, TX75)

Table 1. The reported Beijing blue events and associated meteorological factors

|  | WS(m/s) | T(°C) | TCC(%) | LCC(%) | RH(%) | VIS(Km) |
|---|---|---|---|---|---|---|
| 2008/8/15 (Reported blue) | 1.9 | 25.3 | 61 | 28 | 61 | 30 |
| 2014/11/3 (APEC blue) | 0.8 | 8.4 | 0 | 0 | 57 | 35 |
| 2015/6/11 (Beijing blue) | 3.1 | 53 | 37 | 27.3 | 13.8 | 26.3 |
| 2015/9/3 (Parade blue) | 35.7(instant) | 30.7 | 70 | 29.3 | 42 | 24.4 |

(2) To test the applicability of the variable we choose, referring to the 2013-2014 statistical historical analysis of the daily weather phenomenon record we only have on the Internet, we selected forecasted sunny days in Beijing and compared their meteorological factors. The result is shown in table 2.

Table 2. The reported Beijing sunny days and associated meteorological factors

|  | P(mm) | VIS(Km) | TCC(%) | LCC(%) | SD(h) |
|---|---|---|---|---|---|
| 2011/3/7 | 0 | 30 | 17 | 0 | 10.4 |
| 2011/5/22 | 0 | 25 | 57 | 10 | 12.5 |
| 2011/10/6 | 0 | 30 | 7 | 0 | 10.2 |
| 2012/1/25 | 0 | 20 | 7 | 0 | 8.1 |
| 2012/6/16 | 0 | 30 | 40 | 17 | 11.5 |
| 2012/12/29 | 0 | 30 | 53 | 0 | 7.8 |
| 2013/8/17 | 0 | 30 | 17 | 7 | 12.9 |
| 2013/11/17 | 0 | 35 | 20 | 17 | 8 |

The results show that the AX30, TX70, TX75 account for 100%, 75%, 75% of the forecasted sunny days respectively. According to visibility ranks in Table 3, We choose 'great' visibility(≥20 km) as part of our definition by the table. we also choose the lower quartile visibility (≥15 km) as a comparison. Then, we obtain 6 alternative indexes (shown in table 4).

Table 3. Visibility Ranges in Each Level

| Ranks | range(km) | Grade |
|---|---|---|
| 0 | <0.05 | Worst grade |
| 1 | 0.05-2 | Bad grade |
| 2 | 2-10 | Medium grade |
| 3 | 10-20 | Good grade |
| 4 | >20 | Great grade |

Table 4. The alternative CBDI index definitions

|  | AX30 | TX70 | TX75 |
|---|---|---|---|
| Visibility>=20 km | Blue_1 | Low70_1 | Low75_1 |
| Visibility>=15 km | Blue_2 | Low70_2 | Low75_2 |

(3) The relative humidity will affect visibility by amplifying the light extinction of aerosol particles and gases, bringing the error to human observation(Deng et al., 2011 , Bai et al., 2014 , Chen et al., 2017 , Liu et al., 2017b). Thus, dry visibility (equivalent visibility in dry conditions) is used to reduce the effect of RH. Here, RH values greater than 40% on non-rainy days were converted by formula (1) (Rosenfeld et al., 2007 , Xu et al., 2017)

$$\frac{VIS}{VIS}(dry) = \begin{cases} 0.26 + 0.4285 \times \log 10(100 - RH), 40\% < RH \leq 99\% \\ 0.26 \qquad\qquad , RH > 99\% \end{cases}$$

where RH is in percent and VIS is the observed visibility (km).
It proves that after the adoption of dry visibility, the accuracy rate has increased.

(4) Daily average AQI values in 2012-2014 of Beijing are calculated. The values in March, June, September, and December are selected to test whether the BBI can correspond to grade II air quality (AQ≤100) in four seasons and which definition can contain the dual effects of meteorology and pollution. Also, we lesson from the forecast score used in CMA to precise our analysis (Table 5). The accuracy of prediction is calculated as formula (2)-(4): $TS = \frac{NA+ND}{NA+NB+NC+ND} \times 100\%$ (2), the rate of missing reports (PO) and the empty reports (FAR) are: $PO = \frac{NC}{NA+NC+ND} \times 100\%$ (3), $FAR = \frac{NB}{NA+NB+ND} \times 100\%$ (4).

**Table 5.** Test classification list

| Definition / Observation | CBDI | NOT CBDI |
|---|---|---|
| AQI<=100 | NA | NC |
| AQI> 100 | NB | ND |

(5) The test result of the Beijing station shown in table 6 reveals that blue_2 is best adopted because of the highest accuracy (73.40), the lowest false rate (18.82), and the acceptable range of empty report rate (11.53). And the appearance of the defined Chinese blue index corresponds well to the good day, and it can represent both clear and clean.

**Table 6.** The scores for 6 CBI in Beijing

|  | blue_1 | blue_2 | low70_1 | low70_2 | low75_1 | low75_2 |
|---|---|---|---|---|---|---|
| TS | 71.28 | 73.40 | 70.21 | 72.34 | 60.21 | 71.28 |
| PO | 23.86 | 18.82 | 25.00 | 20.00 | 34.55 | 19.57 |
| FAR | 8.22 | 11.53 | 8.33 | 11.69 | 8.33 | 11.84 |

So finally, we can put out the following BBI definition: no rain (daily precipitation≤0.1mm), low cloud cover ≤AX30, and the observed visibility at 14:00≥15 km. This means no precipitation (no rain or snow), clear sky and great dry visibility. Although containing subjective parts of our index for lacking long-time actual observation records for sunny days and AQI, we try to use pollution data as a conversion to define a more reasonable blue day than only with meteorology data, and it is proved to agree well with the reality.

## 4. Results

4.1 Climatology and long-term trends of Beijing blue days

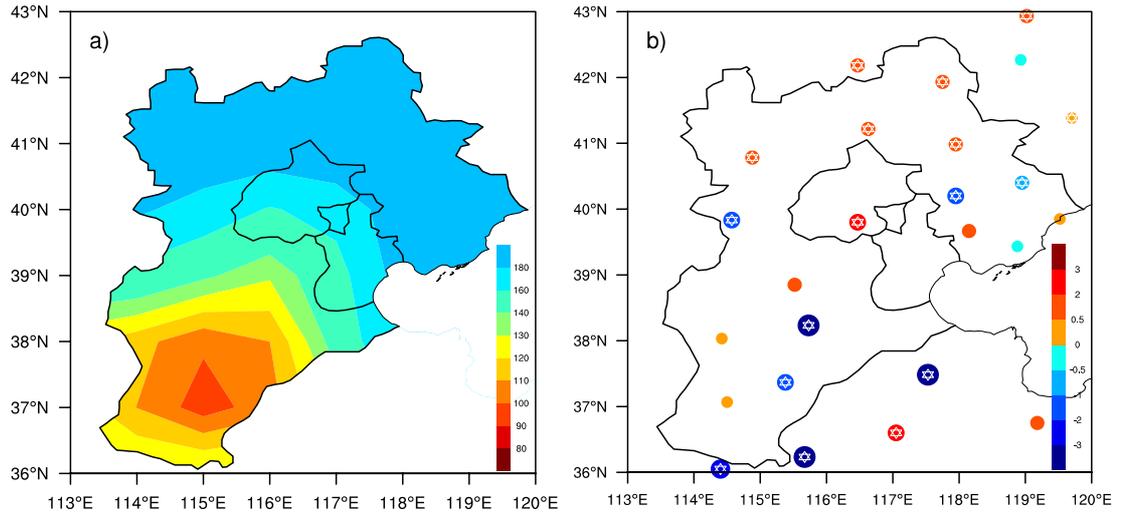

Figure 1. The annual average (a) and the trend coefficients (day/year) (b) of BBI during 1980-2014 over the Beijing-Tianjin-Hebei region.

Figure 1 (a) shows the annual average BBI days from 1980 to 2014 over Beijing-Tianjin-Hebei (BTH). It is clear that the annual average BBI days is decreasing from north to south. Meanwhile, the least annual average BBI occurs in Xing Tai, Hebei province, with 18.86 days per year (d/y), and the largest annual average BBI is located in Mang Ya, Qing Hai province, with 319.65 d/y.

Figure 1 (b) shows the trend coefficient from 1980 to 2014 over BTH. The maximum increasing trend is located in Jin Zhou, Liao Ning province, with 4.83 d/y. and the minimum value occurs in Shen Yang, Liao Ning province, showing a decreasing trend of -4.79 d/y. overall. The BTH annual average BBI days in decreasing at a rate of -0.5 d/10y.

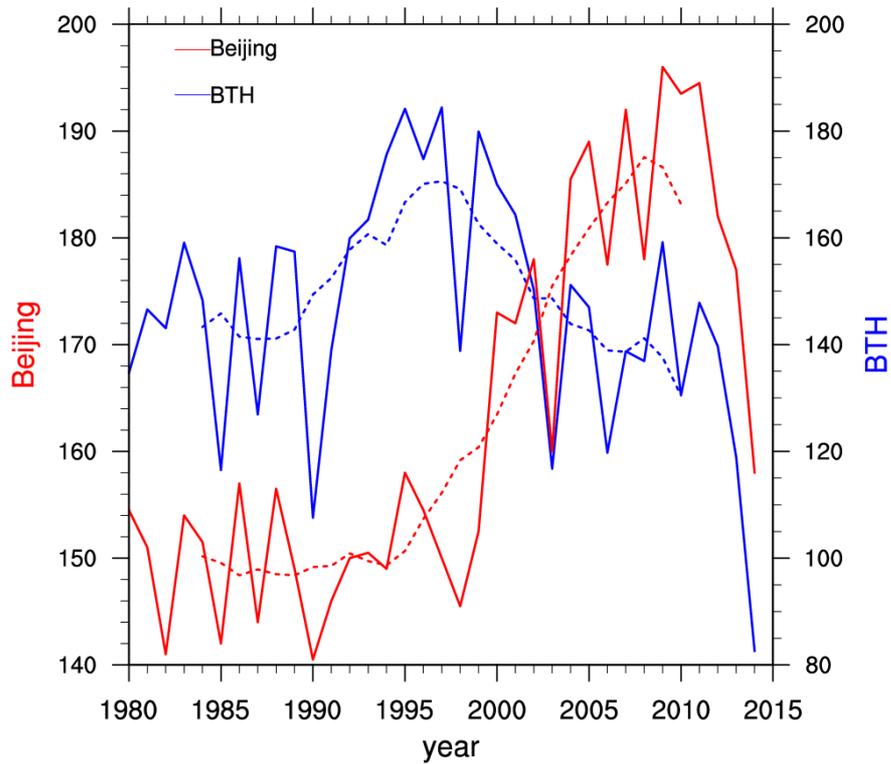

Figure 2. The time series of annual mean of BBI during 1980-2014 in Beijing (red line) and Hebei-Tianjin-Beijing region (blue line). The dotted line represents the 9-year moving average.

The red /blue line in figure 2 represents the time series of annual mean BBI days during 1980-2014 in Beijing/BTH, and the dotted red line represents the 9-year moving average result. BBI days in Beijing show 3 phases——a stable vibration before 1994, a sharp increasing trend in 1994-2008 (with a rate of 5.88 d/y), and an obvious decreasing trend after 2008. The maximum BBI days happens in 2009, with more than 190 d/y, in general, the total trend has a relatively constant increasing rate of 2.66 d/y. meanwhile, the BTH shows a similar change fluctuation before 2001, but an opposite decreasing trend during 2001-2008, then BBI days in BTH show a more distinct decreasing trend from 2019-2014, with more than 6.8 d/y.

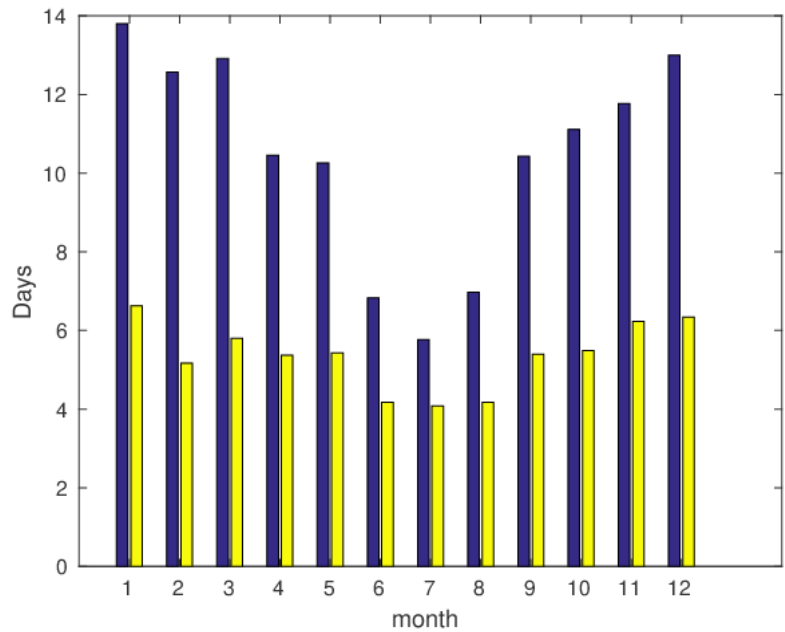

Figure 3. The monthly average and frequency distribution of BBI from the year 1980 to 2014 in Beijing

The monthly average and frequency distribution of BBI from 1980 to 2014 is given in figure 3. The purple line in figure 3 shows the monthly average BBI days in Beijing, while the yellow line in the figure shows the monthly frequency distribution. It can be seen that the maximum monthly average and frequency are in January (13.80 days and 6.63 times). The minimum monthly average frequency happens in July (5.77 days and 4.09 times). From January to July, both monthly average BBI days and frequency distribution are displaying a decreasing trend (-1.37 day/month and -0.36 times/month), and an opposite trend from July to December (1.46 day/month and -0.50 times/month),

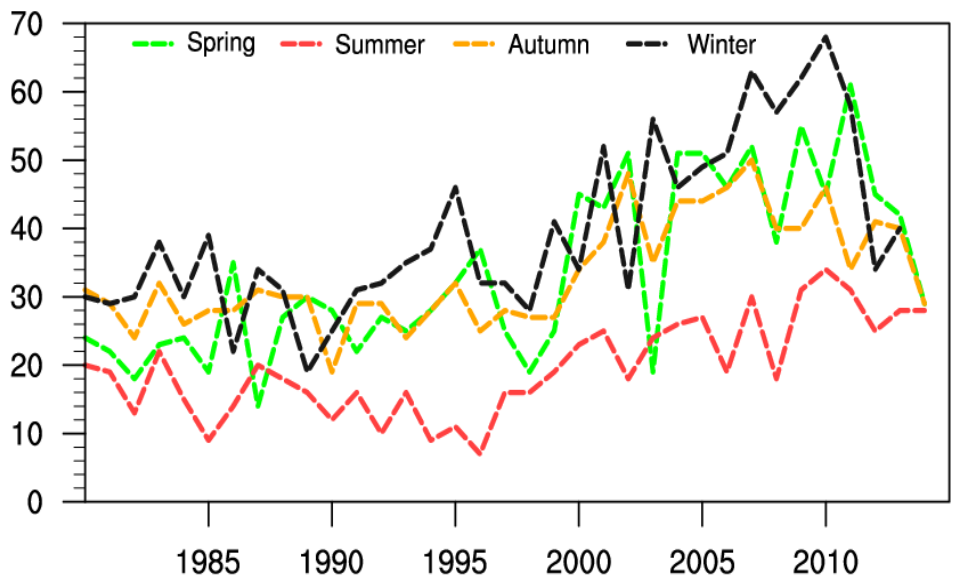

Figure 4. The time series of seasonal distribution of BBI during 1980—2014 in Beijing

Figure 4 shows the time series of seasonal distribution of BBI during 1980—2014 in Beijing. The mean seasonal average BBI days is winter>spring>autumn>summer, with 33.63, 19.57, 33.31, 39.47 days, respectively. The maximum BBI days occurs in winter of 68 days in 2010. And the trend coefficients for four seasons are all showing an increasing trend, with 0.88, 0.46,0.50,0.88 days/season. After all, the increasing trend has translated to decreasing after 2010 and the reasons remain unclear.

4.2 Climatology and long-term trends of continuous Beijing blue days

To further investigate the spatio-temporal distribution of continuous BBI, a continuous Beijing blue event is defined by BBI lasting for more than three days, and continuous blue days in a moth means all blue days lasting more than 3 days. Figure 6 shows the total number and frequency distribution of continuous blue days per month .

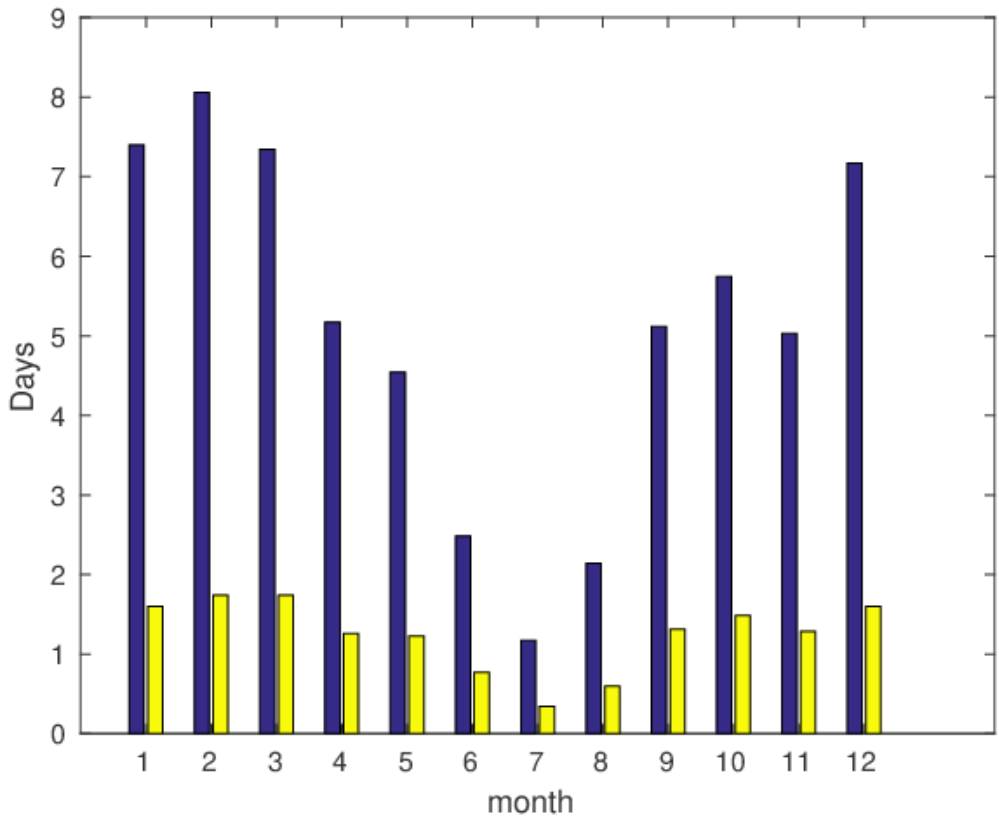

Figure 5. Total number and frequency distribution of continuous blue days per month.

The yellow line in figure 5 represents the frequency of continuous blue days per month and the purple line represents the number of blue days per month. The maximum value is in February (8.05 days 1,74 times). And the minimum value is in July (1.17 days 0.34 times), the frequency of high probability of continuous BBI in a month show small difference and the distribution of frequency corresponds to the monthly average one.

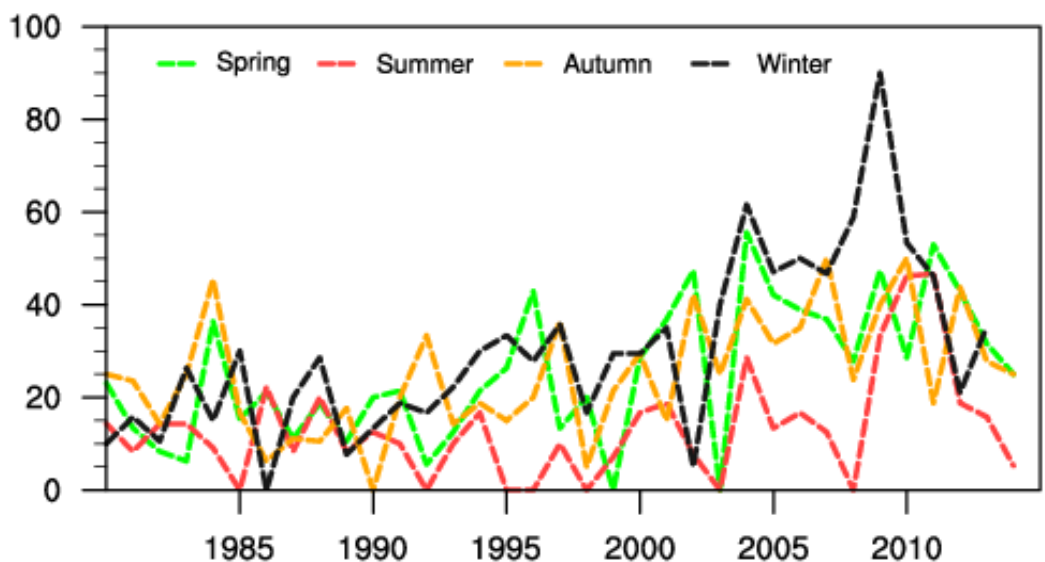

Figure 6. Percentage of continuous blue day frequency in total blue day frequency (%)

Figure 6 shows the percentage of continuous blue day frequency in the total blue day frequency from 1980 to 2014. The total average continuous blue days in four seasons are 25.42 d, 13.30 d, 25.09 d, 30.21 d, respectively, showing a dominant role for continuous BBI days in winter. In 2009, the continuous BBI days account for 90% of total BBI days in winter.

The percentage of continuous BBI in spring is continuously increasing. 1995 is a watershed: the previous summer shows a decreasing trend of volatility, followed by a significant increase, then followed by a rapid decline after 2011. In autumn it shows a decreasing trend before 1990 and then continued to increase after 1990. In winter it shows an increasing trend before 2009 and then decreased rapidly. It increases most rapidly in winter, with a rate of 1.23 d/season, with the same growing trend for total BBI days, the proportion of haze in winter may decrease.

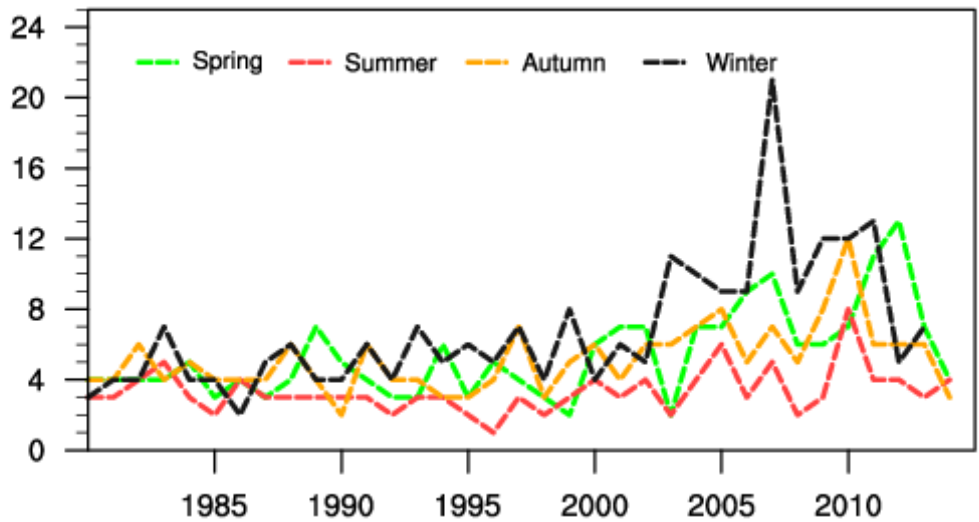

Figure 7. the maximum duration of continuous Beijing blue event during 1980—2014

Figure 7 shows the maximum duration of continuous Beijing blue event for four seasons from 1980 to 2014. The average maximum durations for four seasons are 5.40 d, 3.34 d, 5.17 d, 6.82 d, respectively. Meanwhile, they all represent a growing trend with a rate of winter > spring>autumn>summer. The historical maximum duration lasts up to 21 days and appears in the winter of 2007, after which, the maximum durations begin to decrease.

4.3 Impact of climate conditions on Beijing blue days

Figure 8 shows the long-term trend of annual climatic factors and BBI days during 1980-2014 in Beijing.   The annual mean RH in Beijing is 54.05%, It is clear to find that RH (figure 8a) is negatively correlated to BBI since 1985, and their peaks and valleys are always appeared simultaneously, like 1985,1990,1998,2005 and so on. The RH in Beijing shows an obvious increasing-decreasing two-stage change, with a coefficient -0.55, which passing the 95% confidence level.

Rainless days (figure 8b) in Beijing exceed 265 days a year, with an increasing trend 4.4 d/10y. Temperature (figure 8d) also displays a growing trend of 4.0 d/10y. And they all represent a coordinated variation with BBI for the full year, with a high relation coefficient of 0.47 in the long term, while they also show an opposite change trend after 2010. Wind speed (figure 8c) is weakly correlated to BBI, but the opposite change can be found during 1998-2010.

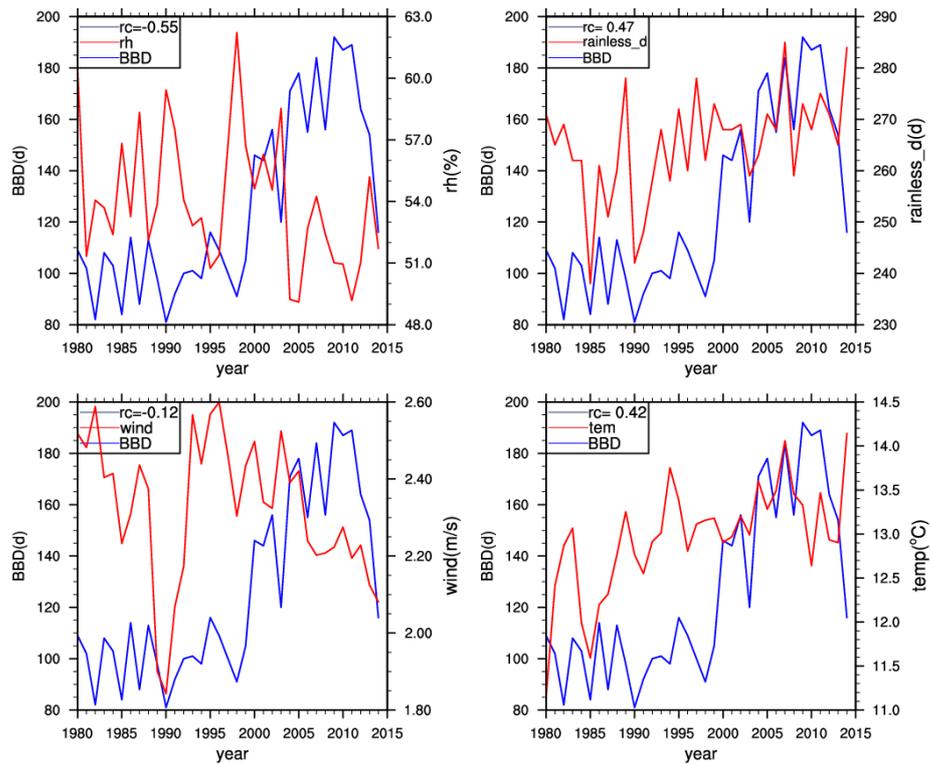

Figure 8. Relationships between climatic factors and BBI over annual scale during 1980—2014

The seasonal average result is also given in table 7. In spring, all factors except for wind speed show a close relationship with BBI, with the highest coefficient -0.63 for RH, rainless days and temperature are positively related to BBI, with the same coefficient 0.38. in summer, RH shows a negative correlation with BBI, this may be affected by more precipitation in summer. With more rain, rainless days will decrease sharply, resulting in a higher RH than other seasons and fewer blue days. Rainless days and temperature show similar results like spring. In autumn and winter, RH is negatively related to BBI, with the correlation coefficient pass -0.40, at the same time, the coefficient between rainless days and BBI pass 0.40 in two seasons, while the temperature and wind speed show a weak link with BBI.

Table 7. Relationships between climatic factors and BBI

| Meteorology | Average | Trend(d/10y) | Correlation coefficient | | | | |
| --- | --- | --- | --- | --- | --- | --- | --- |
| | | | Spring | Summer | Autumn | Winter | Annual |
| RH(%) | 54.05 | 1.1 | -0.63* | 0.60* | -0.47* | -0.40* | -0.55* |
| Rainless days(d) | 265.2 | 4.4 | 0.38* | 0.67* | 0.45* | 0.40* | 0.47* |
| Wind speed(m/s) | 2.33 | -0.1 | 0.02 | 0.06 | -0.16 | -0.07 | -0.12 |
| Temperature(°C) | 12.96 | 0.4 | 0.38* | 0.57* | 0.29 | 0.11 | 0.42* |
| * means the results passing the 95% confidence level | | | | | | | |

## 5. Conclusion and discussions

In this paper, a BBI index, which accuracy is beyond 73%, is defined to analyze the temporal-spatial variations of Beijing blue days. The continuous BBI days and affecting meteorological factors are also discussed.

The results show that Beijing is the most significant accelerating station in the BTH region, whose growth rate is 2.66 d/y. the long term change of BBI shows 3 phases——a stable vibration before 1994, a sharp increasing trend in 1994-2008 (with a rate of 5.88 d/y), and an obvious decreasing trend after 2008.

The greatest seasonal average and the max acceleration rate appears in winter, with a rate of 33.47 d and 0.88 d/season. the minimum acceleration is in summer, which is 0.46 d/season. For continuous Beijing blue days, the maximum and the max acceleration rate appears in winter, too.

We also find that Beijing blue days appears most frequently in January, and least frequently in July.

BBI is more related to relative humidity, temperature, and rainless day on an annual scale, with a coefficient of 0.55, 0.42, and 0.47. But it is different in different seasons. In spring, the dominant factor is RH, in summer it is rainless days, in autumn it is RH and rainless days, while in winter it is RH.

These works would fill up the blank of the blue skies investigation of Beijing and could be helpful to further investigations of the rebuilding of blue China, as well as the three-year action plan to fight air pollution.

**references**

Bai A, Zhong W, Hua L, et al. Analysis on the Variation of Visibility in Chengdu and its Factors of Low Visibility[J]. Environmental Monitoring in China, 2014,30 (2): 21-25.

Cai W, Ke L, Hong L, et al. Weather conditions conducive to Beijing severe haze more frequent under climate change[J]. 2017,7 (4): 257–262.

Chen C, Sun Y L, Xu W Q, et al. Characteristics and sources of submicron aerosols above the urban canopy (260 m) in Beijing, China, during the 2014 APEC summit[J]. Atmospheric Chemistry and Physics, 2015,15 (22): 12879-12895.

Chen Z Y, Cai J, Gao B B, et al. Detecting the causality influence of individual meteorological factors on local PM2.5 concentration in the Jing-Jin-Ji region[J]. Scientific Reports, 2017,7

Deng J J, Wang T J, Jiang Z Q, et al. Characterization of visibility and its affecting factors over Nanjing, China[J]. Atmospheric Research, 2011,101 (3): 681-691.

Gao M, Liu Z R, Wang Y S, et al. Distinguishing the roles of meteorology, emission control measures, regional transport, and co-benefits of reduced aerosol feedbacks in "APEC Blue"[J]. Atmospheric Environment, 2017,167: 476-486.

Guo J P, He J, Liu H L, et al. Impact of various emission control schemes on air quality using WRF-Chem during APEC China 2014[J]. Atmospheric Environment, 2016,140: 311-319.

Huang K, Zhang X Y, Lin Y F The "APEC Blue" phenomenon: Regional emission control effects observed from space[J]. Atmospheric Research, 2015,164: 65-75.

Kang Z, Gui H, Wang J, et al. Characteristics and cause of the parade blue in Beijing 2015[J]. China Environmental Science, 2016,36 (11): 3227-3236.

Li H, Zhang Q, Duan F, et al. The "Parade Blue": effects of short-term emission control on aerosol chemistry[J]. Faraday Discussions, 2016,189: 317-335.

Li X, Qiao Y, Zhuc J, et al. The "APEC blue" endeavor: Causal effects of air pollution regulation on air quality in China[J]. Journal of Cleaner Production, 2017,168: 1381-1388.

Lin C, Guangyu S H I, Shiguang Q I N, et al. Analysis of Physical Properties of Aerosols during the 2008 Beijing Olympic Games[J]. Climatic and Environmental Research, 2009,14 (6): 665-672.

Lin J, Donkelaar A V, Xin J, et al. Clear-sky aerosol optical depth over East China estimated from visibility measurements and chemical transport modeling[J]. 2014,95 (1): 258–267.

Liu H, Liu C, Xie Z, et al. A paradox for air pollution controlling in China revealed by "APEC Blue" and "Parade Blue"[J]. Scientific Reports, 2016,6

Liu H L, He J, Guo J P, et al. The blue skies in Beijing during APEC 2014: A quantitative assessment of emission control efficiency and meteorological influence[J]. Atmospheric Environment, 2017a,167: 235-244.

Liu H M, Fang C L, Zhang X L, et al. The effect of natural and anthropogenic factors on haze pollution in Chinese cities: A spatial econometrics approach[J]. Journal of Cleaner Production, 2017b,165: 323-333.

Ramanathan V, Ramana M V Persistent, widespread, and strongly absorbing haze over the Himalayan foothills and the Indo-Gangetic Plains[J]. Pure and Applied Geophysics, 2005,162 (8-9): 1609-1626.

Rosenfeld D, Dai J, Yu X, et al. Inverse relations between amounts of air pollution and orographic precipitation[J]. Science, 2007,315 (5817): 1396-1398.

Shen J L, Tang A H, Liu X J, et al. Impacts of Pollution Controls on Air Quality in Beijing during the 2008 Olympic Games[J]. Journal of Environmental Quality, 2011,40 (1): 37-45.

Shen X J, Sun J Y, Zhang X Y, et al. Characterization of submicron aerosols and effect on visibility during a severe haze-fog episode in Yangtze River Delta, China[J]. Atmospheric Environment, 2015,120: 307-316.

Streets D G, Fu J S, Jang C J, et al. Air quality during the 2008 Beijing Olympic Games[J]. Atmospheric Environment, 2007,41 (3): 480-492.

Su B D, Zhan M J, Zhai J Q, et al. Spatio-temporal variation of haze days and atmospheric circulation pattern in China (1961-2013)[J]. Quaternary International, 2015,380: 14-21.

Wang G, Cheng S, Wei W, et al. Characteristics and emission-reduction measures evaluation of PM2.5 during the two major events: APEC and Parade[J]. Science of the Total

Environment, 2017,595: 81-92.

Wang S, Huang G, Lin J, et al. Chinese Blue Days: A novel index and spatio-temporal variations[J]. 2019,14 (7)

Wang S, Zhao M, Xing J, et al. Quantifying the Air Pollutants Emission Reduction during the 2008 Olympic Games in Beijing[J]. Environmental Science & Technology, 2010,44 (7): 2490-2496.

Wang W T, Primbs T, Tao S, et al. Atmospheric Particulate Matter Pollution during the 2008 Beijing Olympics[J]. Environmental Science & Technology, 2009,43 (14): 5314-5320.

Xu X D, Guo X L, Zhao T L, et al. Are precipitation anomalies associated with aerosol variations over eastern China?[J]. Atmospheric Chemistry and Physics, 2017,17 (12): 8011-8019.

Xue Y, Wang Y, Li X, et al. Multi-dimension apportionment of clean air "parade blue" phenomenon in Beijing[J]. Journal of Environmental Sciences, 2018,65: 29-42.

Zhang X Y, Wang Y Q, Niu T, et al. Atmospheric aerosol compositions in China: spatial/temporal variability, chemical signature, regional haze distribution and comparisons with global aerosols (vol 12, pg 779, 2012)[J]. Atmospheric Chemistry and Physics, 2012,12 (14): 6273-6273.

Zhang Z, Zhang X, Gong D, et al. Possible influence of atmospheric circulations on winter haze pollution in the Beijing-Tianjin-Hebei region, northern China[J]. Atmospheric Chemistry and Physics, 2016,16 (2): 561-571.